\documentclass[11pt,a4paper]{article}
\usepackage{jheppub}
\usepackage{latexsym, graphicx}
\usepackage{amssymb}
\usepackage{amsmath}
\usepackage{subfig}

\def\e{\epsilon}

\title{Boundary Effects on Quantum Entanglement and its Dynamics in a Detector-Field System}

\author[a]{Rong Zhou,}
\author[b]{Ryan O. Behunin,}
\author[c,1]{Shih-Yuin Lin,}
\author[a]{and B. L. Hu}

\affiliation[a]{Joint Quantum Institute and Maryland Center for Fundamental
Physics, University of Maryland, College Park, Maryland 20742, USA}

\affiliation[b]{Center for Nonlinear Studies and Theoretical Division, 
Los Alamos National Laboratory,\\ Los Alamos, New Mexico 87545, USA}

\affiliation[c]{Department of Physics, National Changhua University of Education,\\
Changhua 50007, Taiwan \note{Corresponding author.}}

\emailAdd{zhour@umd.edu}
\emailAdd{rbehunin@lanl.gov}
\emailAdd{sylin@cc.ncue.edu.tw}
\emailAdd{blhu@umd.edu}

\abstract{
In this paper we analyze an exactly solvable model consisting of an inertial Unruh-DeWitt detector
which interacts linearly with a massless quantum field in Minkowski spacetime with a perfectly reflecting flat plane boundary.
Firstly a set of coupled equations for the detector's and the field's Heisenberg operators are derived.
Then we introduce the linear entropy as a measure of entanglement between the detector and the quantum field under mirror reflection,
and solve the early-time detector-field entanglement dynamics.
After coarse-graining the field, the dynamics of the detector's internal degree of freedom is described by
a quantum Langevin equation, where the dissipation and noise kernels respectively correspond to the retarded Green's functions and
Hadamard elementary functions of the free quantum field in a half space.
At late times when the combined system is in a stationary state, we obtain exact expressions for the detector's
covariance matrix and show that the detector-field entanglement decreases
for smaller separation between the detector and the mirror.
We explain the behavior of detector-field entanglement qualitatively with the help of a detector's mirror image, compare
them with the case of two real detectors and explain the differences.
}
\keywords{Boundary quantum field theory, quantum dissipative system, relativistic quantum information.}

\begin{document}
\maketitle

\section{Introduction}

The effects on a quantum field due to the presence of boundaries (such as a mirror or a dielectric plate) 
has been studied a long time ago, as in the famous Casimir effect \cite{Casimir}. 
Moving mirrors in a quantum field have been used as analog models \cite{DavFul} of Hawking and Unruh effects \cite{Hawk75, Unr76}.  
Quantum field theory in spacetimes with nontrivial topology \cite{QFTBounTop} has also been studied since the seventies. 
Today these subject matters fall under the broadened scope known as quantum field theory under external conditions \cite{QFText}.

Quantum entanglement is, according to Schr\"odinger, ``{\it the} characteristic trait of quantum mechanics, 
the one that enforces its entire departure from classical line of thought" \cite{Schr35}.  
The nature and behavior, even the definition for more than two parties, of entanglement has attracted much attention 
in the last two decades for both theoretical and practical reasons, the former because it is the cornerstone issue 
at the foundation of quantum mechanics and the latter connected with the rapid development of quantum information sciences.

Quantum entanglement of systems composed of detectors \footnote{By detector 
we refer to a physical entity with some internal degrees of freedom, e.g. an atom, an electron with spin, 
or an oscillator. We will be working with the Unruh-DeWitt (UD) detector specifically \cite{Unr76, DeW79}.} 
and mirrors, coexisting with and/or mediated by a quantum field is also receiving increased attention as an active research branch 
in a newly emergent field known as relativistic quantum information \cite{RQI} -- 
relativistic because the effects of the quantum field is highlighted where traditional treatments based on nonrelativistic quantum mechanics 
prove inadequate. (For a recent review of problems in RQI involving UD detectors see, e.g., \cite{HLL}). 
An important theoretical issue which quantum field theory brings forth in contrast to quantum mechanics is quantum nonlocality,  
a vivid illustrative example of this is given in \cite{LinHuOLCE}, 
where entanglement between two causally disconnected objects can be created by their local couplings with a common quantum field.

Amongst theoretical problems on quantum entanglement of current experimental interest \cite{QIPspecial} we mention two aspects: 
1) Time evolution, or entanglement dynamics:
Even in the simplest system of two two-level (2LA) atoms interacting via a common quantum field already one sees 
a diverse spectrum of interesting behavior, ranging from sudden death \cite{YuEberly}, touch of death, revival, to staying always alive, 
and features such as dynamical generation, protection, and transfer of entanglement between subsystems \cite{SCH1}. 
Model studies of oscillator-field entanglement \cite{PazRoncaglia} with experimental ventures have been carried out \cite{Sabrina} 
as well as mirror-field entanglement in conjunction with measurements in LIGO detectors \cite{Miao}, 
the latter also serving as preparatory studies for quantum superpositions of mirrors \cite{Marshall} in macroscopic quantum phenomena. 
2) Spatial dependence: 
Entanglement not only changes in time but also depends on spatial separation,  as shown in model studies for 2LAs \cite{ASH,FCAH_2atom,SCH2} 
and harmonic oscillators \cite{LH_2IHO} interacting with a common quantum field. 
Practical application of these studies span from  the design of quantum gates to quantum teleportation \cite{SLCH}. 

In this paper we study the model consisting of an inertial UD detector which interacts linearly with a massless quantum field, 
the latter being restricted to a half space by a perfectly reflecting boundary,
or equivalently, a perfect mirror. The presence of the mirrors is known to influence radiative properties of physical systems, 
such as the rate of spontaneous emission of excited atoms \cite{Purcell}. 
Here we probe such systems at a deeper level, asking how the mirror affects
the delicate property of entanglement between the detector and the quantum field at zero temperature.

The effect of the mirror can be understood from two perspectives. From a purely technical consideration,
the presence of the mirror breaks Poincar\'e symmetry and thus modifies the spectral density of the field,
which accordingly alters the entanglement and its dynamics of the detector-field system. 
From physical considerations one can exploit the symmetry in the set-up and see if this problem of
a detector interacting with a mirror-modified field can be solved more simply with the use of a mirror image,
as is familiar in classical electromagnetics. From this perspective, one may even ask whether a detector can be entangled
with its own mirror image, leading to the notion of self-entanglement ( \textit{autangle} or \textit{ipso-tangle} ).
We will pursue this problem in a cautious  manner, i.e., carry out the full calculation without assuming any mirror image;
only after we obtain the results will we then ask if one could interpret them in a mirror image way and answer the query,
``Can one get entangled with one's mirror image?"

The paper is organized as follows: In Section \ref{setup} we set up the
problem and derive a formal set of equations for the Heisenberg operators
of the detector's internal degrees of freedom and the quantum field. We
impose the boundary conditions introduced by the presence of the mirror,
then calculate the retarded Green's functions and Hadamard elementary functions of the altered field configurations.
In Section \ref{early} we introduce the linear entropy as a measure of quantum entanglement
and calculate the early-time dynamics of entanglement between the detector and
the field in our setup. 
We show how the entanglement between the detector and field in
a half space under mirror reflection evolves in time, and how it depends on the distance
between the detector and the mirror.
In Section \ref{latetime} we derive a quantum Langevin equation for the detector's internal degree of freedom 
including the back-reaction of the altered field. We seek 
 late-time solutions to this equation
and calculate the covariance matrix of the detector's canonical variables
at late times when the combined system is in a stationary state.

Details are contained in Appendix \ref{LateCV}.
Finally in Section \ref{discuss} 
we explain in what sense can one describe this situation in terms
of entanglement of the detector with its mirror image and why it decreases
as the detector moves closer to the mirror, a somewhat counterintuitive
finding. We also compare this situation with the case of two inertial
detectors in free space (calculations placed in Appendix \ref{w2IHO}) and explain
their physical differences.

\section{Unruh-DeWitt Detector in Half Space under Mirror Boundary Condition}
\label{setup}

The gist of the matter for this paper is to quantify the change of entanglement between
the detector and the field because of the presence of a mirror. To achieve this we first derive the Heisenberg equations of motion
for both the detector's internal degree of freedom and the field under the boundary condition of the mirror.
After coarse graining the field, we obtain the reduced dynamics of the detector
which already contains the back-reaction of the field on the detector.
Then by solving the reduced dynamics of the detector one can obtain its time-dependent covariance matrix. This covariance matrix describes the mixed state of the detector's
internal degree of freedom, according to which the detector-field entanglement can be quantified because the combined detector-field system is in a pure state.

\subsection{Dynamics of Detector-Field System with Mirror}

Consider an UD detector located at position ${\bf x}_q^{}$ at a vertical distance of $L/2$ from the mirror plane at $x_3=0$.
The action of the total system, which describes the detector with internal degree of freedom $Q$ interacting linearly
with a massless scalar field $\Phi$ in a half space defined by $x_{3}>0$
with coupling constant $\lambda_q$, is given by
\begin{equation}
    S
    =\frac{1}{2}M_{q}\int dt\ (\dot{Q}^{2}-\Omega_0^{2}Q^{2})+
    \frac{1}{2}\int dt\int_{x_{3}>0}d^{3}x \partial_{\mu}\Phi\partial^{\mu}\Phi+\lambda_{q}\int dt\ Q(t)\Phi({\bf x}_q^{},t),
\label{eq:Action_Half}
\end{equation}
where $Q$ acts like a harmonic oscillator with mass $M_q$ and bare natural frequency $\Omega_0$. 
Due to the linearity of our system the equations of motion for Heisenberg
operators (carrying hats), 
\begin{eqnarray}
& & M_{q}\ddot{\hat{Q}}(t)+M_{q}\Omega^{2}{\hat{Q}}(t)= \lambda_{q} 
    \hat{\Phi}({\bf x}_q^{},t),
\label{AtomEqn}\\
& & \Box{\hat{\Phi}}({\bf x},t)= \lambda_{q}\delta^{3}({\bf x}-{\bf x}_q^{}){\hat{Q}}(t), \label{FieldEqn} \\
& & 
{\hat{\Phi}}({\bf x}_{\|},x_{3}=0,t)=0,\label{DBC}
\end{eqnarray}
have the same form as the equations for the corresponding classical variables.
Equation (\ref{DBC}) specifies the Dirichlet boundary
conditions imposed on the field where the mirror surface is located,
namely, in the $x_{3}=0$ plane and ${\bf x}_{\|}=(x_1,x_{2})$. To
obtain the back-reaction of the field on the detector we first solve (\ref{FieldEqn})
and then plug the solution into (\ref{AtomEqn}). The solution for
${\hat{\Phi}}({\bf x},t)$ is given by
\begin{equation}
  {\hat{\Phi}}({\bf x},t)={\hat{\Phi}}_{0}({\bf x},t)+\lambda_{q}\int_{t_{i}}^{t}dt'G_{ret}^{\Phi}(t,{\bf x};t',
  {\bf   x}_q^{}){\hat{Q}}(t')\label{FieldSoln}
\end{equation}
where ${\hat{\Phi}}_{0}({\bf x},t)$ is the homogeneous solution
to equation (\ref{FieldEqn}) which describes the dynamics of the
source-free field (without $Q$) in the presence of the mirror and
$G_{ret}^{\Phi}({\bf x},t;{\bf y},t^{\prime})$ is the retarded Green's
function for the field. Plugging the solution for the field operator
into the equation of motion for the detector we obtain the following
equation governing the dynamics of the detector with the effects of the
field already incorporated,
\begin{equation}
  M_{q}\ddot{\hat{Q}}(t)+M_{q}\Omega^{2}{\hat{Q}}(t)-\lambda_{q}^{2}\int_{t_{i}}^{t}dt'G_{ret}^{\Phi}(t,{\bf x}_q^{};t',
    {\bf x}_q^{}){\hat{Q}}(t')=\lambda_{q}{\hat{\Phi}}_{0}({\bf x}_q^{},t).
\label{FI-AtomEqn}
\end{equation}
This is what we mean by the `field-influenced' dynamics of the detector.
The term containing the
retarded Green's function describes how the detector transfers energy
to the field, as will be explained in further detail later,  
and the right hand side acts similarly to a Langevin
forcing term describing how quantum field fluctuations drive the detector.

\subsection{Green's Functions of Free Field in Half-space}

In order to solve the equation of motion for the detector's internal degree of freedom (\ref{FI-AtomEqn}),
we need to know the evolution of free field in a half space. The free field operator $\hat{\Phi}_{0}$
which satisfies the Klein-Gordon equation and vanishes in the $x_{3}=0$
plane can be decomposed into normal modes:
\begin{equation}
  {\hat{\Phi}}_{0}({\bf x},t)=\int_{k_{3}>0}d^{3}k\sqrt{\frac{1}{4\pi^{3}\omega}}
  e^{-i\omega t+i{\bf k}_{\|}\cdot {\bf x}_{\|}}\sin k_{3}x_{3}{\hat{b}}_{{\bf k}}+H.c. 
\end{equation}
where $\omega=|{\bf k}|$.
Notice that the normalization factor is different compared to the case of free space.

Writing $\hat{\Pi}_{0}({\bf x}',t)=\dot{\hat{\Phi}}_{0}({\bf x}',t)$
and demanding that the equal-time commutation relations are satisfied in half-space,
that is,
\begin{equation}
  [\hat{\Phi}_{0}({\bf x},t),\hat{\Pi}_{0}({\bf x}',t)]=i\delta^{(3)}({\bf x}-{\bf x}'),\label{CommRel}
\end{equation}
we can infer that the mode expansion coefficients can be identified
with the creation and annihilation operators
\begin{equation}
  [{\hat{b}}_{{\bf k}},{\hat{b}}_{{\bf k}}^{\dag}]=\delta_{{\bf k},{\bf k}'},
  \quad[{\hat{b}}_{{\bf k}},{\hat{b}}_{{\bf k}}]=0,
  \quad[{\hat{b}}_{{\bf k}}^{\dag},{\hat{b}}_{{\bf k}}^{\dag}]=0.
\end{equation}
In the above commutators the vectors are restricted to a half space, i.e. $x_3>0$ and $k_3>0$.

With the expression for the free field operator in hand we now calculate
the retarded Green's function and Hadamard elementary function. The retarded Green's function quantifies
the field sourced by the detector and describes how energy is dissipated
from the detector into the field:
\begin{eqnarray}
  G_{ret}^{\Phi}({\bf x},t;{\bf y},t^{\prime}) & \equiv & i\theta(t-t^{\prime})[\hat{\Phi}_{0}({\bf x},t),\hat{\Phi}_{0}({\bf y},t')]\nonumber \\
  &=&\theta(t-t^{\prime})\frac{1}{2\pi^{3}}\int_{k_{3}>0}d^{3}k\frac{1}{\omega}\sin(\omega(t-t^{\prime}))e^{i{\bf k}_{\parallel}
    \cdot({\bf x}_{\parallel}-{\bf y}_{\parallel})}\sin(k_{3}x_{3})\sin(k_{3}y_{3})\nonumber\\
  &=&\theta(t-t^{\prime})\int_{0}^{\infty}d\omega\sin(\omega(t-t^{\prime}))\cdot I(\omega;{\bf x},{\bf y}),
\end{eqnarray}

where the spectral density $I(\omega;{\bf x},{\bf y})$ is given by
\begin{equation}
  I(\omega;{\bf x},{\bf y})=\frac{\omega}{2\pi^{3}}\int_{0}^{\pi/2}d\theta\int_{0}^{2\pi}d\phi\ e^{i{\bf k}_{\|}\cdot({\bf x}_{\parallel}
  -{\bf y}_{\parallel})}\sin(\omega\cos\theta x_{3})\sin(\omega\cos\theta y_{3}).
\label{SpectralDensity}
\end{equation}
The Hadamard elementary function, on the other hand,
describes the quantum fluctuations of the field. It is given formally by the anti-commutator of the field operator:
\begin{eqnarray}
  G_{H}^{\Phi}({\bf x},t;{\bf y},t^{\prime}) &\equiv& \langle\{\hat{\Phi}_{0}({\bf x},t),\hat{\Phi}_{0}({\bf y},t')\}\rangle\nonumber \\
  &=&\frac{1}{2\pi^{3}}\int_{k_{3}>0}d^{3}k\frac{1}{\omega}\cos(\omega(t-t^{\prime}))e^{i{\bf k}_{\parallel}\cdot({\bf x}_{\parallel}
    -{\bf y}_{\parallel})}\sin(k_{3}x_{3})\sin(k_{3}y_{3})\nonumber\\
  &=&\int_{0}^{\infty}d\omega\cos(\omega(t-t^{\prime}))\cdot I(\omega;{\bf x},{\bf y}),
\end{eqnarray}
which is derived from the same spectral density (\ref{SpectralDensity}).

\section{Early Time Dynamics of Detector-Field Entanglement}
\label{early}

Understanding how entanglement  in various types of quantum systems evolves in time is of basic importance for 
the purpose of quantum information processing \cite{QIPspecial}
and can provide new insight into certain foundational issues involving apparent information loss and unitarity violation.
For example, the dynamics of entanglement between an accelerating UD detector and a massless field in 3+1 dimensional 
Minkowski spacetime has been shown to have implications for the issue of black hole information \cite{LinHu08}.

In \cite{LH_2IHO} the dynamics of the entanglement between two inertial detectors
with identical couplings to a common massless scalar field was investigated
and it exhibited several interesting features.
There are roughly three stages of evolution
if the two detectors are properly separated
and weakly coupled to the field. The first stage 
is at early times, up to $t\approx O(1/\lambda_q^2)$ when accumulated effects of 
field-mediated mutual influences between the detectors,
which manifest themselves in the field's retarded propogators,
are still weak and the reduced dynamics of the detectors are dominated by the influences of vacuum fluctuations of the field.
As the effects of field-mediated mutual influences between the detectors gradually gain strength, the system will enter an intermediate stage
in which the detectors' reduced dynamics become quite complicated.
In the late-time limit, or the final stage,  
the detector approaches a stationary state in time.

In particular, in \cite{LH_2IHO} it was shown that at early times,
after causal contact has been established between two detectors,
an oscillatory pattern of entanglement emerges which varies with the distance between the detectors.
This is not due to mutual influences since in the weak coupling limit the mutual influence effect
is always small 
compared to effect of vacuum fluctuations.

For our model, we expect a similar partition of the history in three stages.
Here in this section we study the early-time behavior of our model, assuming
that initially the detector is in its ground state uncorrelated with the field, 
which is also in its ground state under an external Dirichlet boundary condition. 
We will show that the detector-field entanglement
develops a spatial oscillatory pattern characterized by the renormalized frequency of the detector
$\Omega_r$ (we adopt the convention $c=1$).
Moreover, the growth rate of the detector-field entanglement
also oscillates as a function of the detector-boundary spacing.

\subsection{Measure of Detector-Field Entanglement }

In our setup, the detector and the field together form a closed quantum
system which undergoes unitary evolution. Since the initial state
of the entire system is Gaussian and the action is quadratic, the reduced
quantum state of the detector at any time will remain Gaussian. For such systems the amount
of bipartite entanglement between the detector and the field can be quantified
by the linear entropy of the detector's reduced density matrix \cite{OneModePurity},
which takes on values larger than $0$ when the detector and field are entangled.
Here the linear entropy is defined as
$S_{L}=1-\mathcal{P}$,
where $\mathcal{P}\equiv Tr\hat{\rho}_{\rm a}^{2}$ is the purity of
the detector's reduced quantum state. 
For our general mixed Gaussian state, the purity is given by
$\mathcal{P}=
1/(2\sqrt{\det \mathbf{V}})$
where $\mathbf{V}$ is the detector's covariance matrix defined as $\mathbf{V}_{ij}=\langle\hat{O}_{i},\hat{O}_{j}\rangle$, in which $\langle\hat{O}_{i},\hat{O}_{j}\rangle \equiv Tr(\hat{\rho}_{\rm a}\cdot\{\hat{O}_{i},\hat{O}_{j}\})/2$,
and $\boldsymbol{\hat{O}}=(\hat{Q},\hat{P})$.
Mathematically we know that $\mathcal{P}$ is always less than or equal to 1.
Smaller purity means the reduced density matrix of the detector
is less pure, and correspondingly the detector is more entangled with the field.

\subsection{Mode Decomposition and Exact Solutions for Mode Functions}

We perform the following mode decompositions for Heisenberg operators
of the detector $\hat{Q}(t)$ and the field $\hat{\Phi}(x)$ with
respect to the creation and annihilation operators of the detector
$\hat{a},\hat{a}^{\dagger}$ and those of the field $\hat{b},\hat{b}^{\dagger}$:
\begin{eqnarray}
  \hat{Q}(t) &=& \sqrt{\frac{\hbar}{2\Omega_{r}}}\left[q_{\rm a}(t)\hat{a}+q_{\rm a}^{*}(t)\hat{a}^{\dagger}\right]+
    \int_{k_{3}>0}\frac{d^{3}k}{\sqrt{2\pi^{3}}}\sqrt{\frac{\hbar}{2\omega}}\left[q_{+}(t,{\bf k})\hat{b}_{{\bf k}}+
    q_{-}(t,{\bf k})\hat{b}_{{\bf k}}^{\dagger}\right],\label{OExpan}\\
  \hat{\Phi}(x) &=& \sqrt{\frac{\hbar}{2\Omega_{r}}}\left[f_{\rm a}(x)\hat{a}+f_{\rm a}^{*}(x)\hat{a}^{\dagger}\right]+
    \int_{k_{3}>0}\frac{d^{3}k}{\sqrt{2\pi^{3}}}\sqrt{\frac{\hbar}{2\omega}}\left[f_{+}(x;{\bf {\bf k}})\hat{b}_{{\bf k}}+
    f_{-}(x;{\bf {\bf k}})\hat{b}_{{\bf k}}^{\dagger}\right],\label{PhiExpan}
\end{eqnarray}
where $\Omega_r$ is the renormalized internal frequency of the detector.
From here on subscript `a' is always
associated with the detector's internal degree of freedom.

By plugging \ref{OExpan} and \ref{PhiExpan} into \ref{AtomEqn} and \ref{FieldEqn}, respectively, we find the following equations of motion for the mode functions
\begin{eqnarray}
  \left(\partial_{t}^{2}+2\gamma_q^{}\partial_{t}+\Omega_{r}^{2}\right)q_{\rm a}(t) &=&
    -\frac{2\gamma_q^{}}{L}\theta(t-L)q_{\rm a}(t-L),\label{EOMqq}\\
  \left(\partial_{t}^{2}+2\gamma_q^{}\partial_{t}+\Omega_{r}^{2}\right)q_{+}(t,{\bf k}) &= &
    -\frac{2\gamma_q^{}}{L}\theta(t-L)q_{+}(t-L,{\bf k})+{\lambda_{q}\over M_q}f_{0+}(t,{\bf x}_q^{};{\bf k}),\label{EOMq+}
\end{eqnarray}
in which $f_{0+}(t,{\bf x}_q^{};{\bf k})=e^{-i\omega t+i{\bf k}_{\parallel}\cdot{\bf x}_{\parallel}}\sin k_{3}x_{3}$
and $\gamma_q^{}\equiv \lambda_q^2/(8\pi M_q)$. Since we have assumed the detector's position to be
${\bf x}_q^{}(t)=\bar{\bf x}_q^{} \equiv (0,0,L/2)$, we have $f_{0+}(\bar{\bf x}_q^{},t;{\bf k})=e^{-i\omega t}\sin\frac{k_{3}L}{2}$.
According to our initial condition, the solutions have to satisfy
the initial conditions 
$f_{+}(0,{\bf x};{\bf k})= e^{i {\bf k}_{\parallel}\cdot{\bf x}_{\parallel}}\sin k_{3}x_{3}$,
$\partial_{t}f_{+}(0, {\bf x};{\bf k})=-i\omega e^{i{\bf k}_{\parallel}\cdot{\bf x}_{\parallel}}\sin k_{3}x_{3}$,
$q_{\rm a}(0)=1$, $\partial_{t}q_{\rm a}(0)=-i\Omega_{r}$,
and $f_{\rm a}(0,{\bf x})=\partial_{t}f_{\rm a}(0,{\bf x})=q_{+}(0;{\bf k})=\partial_{t}q_{+}(0;{\bf k})=0$.

The solution to (\ref{EOMqq}) can be written as the expansion
\begin{equation}
q_{a}(t)=\sum_{n=0}q_{a}^{(n)}(t),\label{qexpan}\end{equation}
with the $n$-th order retarded influences $q_{a}^{(n)}$ 
by the $(n-1)$-th order backreaction sourced from the detector then reflected by the mirror:
\begin{eqnarray}
q_{\rm a}^{(0)}(t) &=& q_{\rm a}^{(h)}(t),\\
q_{\rm a}^{(n)}(t) &=& \int d\tau_{1}G_{r}(t,\tau_{1})\left(-\frac{2\gamma_q^{}}{L}\right)\theta(t-L)\times\nonumber\\
 & &...\int d\tau_{n}G_{r}(\tau_{n-1}-\tau_{n})\left(-\frac{2\gamma_q^{}}{L}\right)\theta(\tau_{n}-nL)q_{\rm a}^{(h)}(\tau_{n}-nL).
\end{eqnarray}
The expansion of $q_a(t)$ is truncated at $n=N$ with the minimal $N$ satisfying $NL>t$, 
because the least time it takes for the $n$-th order field-mediated influence to affect the detector is $nL$.
Here $G_{r}(t,\tau)$ is the retarded Green's function which satisfies
$(\partial_{t}^{2}+2\gamma_q^{}\partial_{t}+\Omega_{r}^{2})G_{r}(t,\tau)=\delta(t-\tau)$ and
\begin{equation}
q_{\rm a}^{(h)}(t)=\frac{1}{2}\left(1+\frac{\Omega_{r}+i\gamma_q^{}}{\Omega}\right)e^{-\gamma_q^{} t-i\Omega t}+\frac{1}{2}\left(1-\frac{\Omega_{r}+i\gamma_q^{}}{\Omega}\right)e^{-\gamma_q^{} t+i\Omega t}
\end{equation}
is the homogeneous solution satisfying the initial conditions.

The solution to (\ref{EOMq+}) can be written in a similar fashion as above, also truncated at $nL>t$,
\begin{equation}
q_{+}(t,{\bf k})=\sum_{n=0}q_{+}^{(n)}(t,{\bf k}),\label{pexpan}
\end{equation}
where
\begin{eqnarray}
q_{+}^{(0)}(t,{\bf k}) &=& \int d\tau_{1}G_{r}(t,\tau_{1})\lambda_{q}f_{0+}(\tau_{1},\bar{\bf x}_q^{};{\bf k})\nonumber\\
 &=& \frac{\lambda_{q}}{M_q\Omega}\left(\sin\frac{k_{3}L}{2}\right)
   \left[(M_{1}-M_{2})e^{-i\omega t}+\left(M_{2}e^{i\Omega t}-M_{1}e^{-i\Omega t}\right)e^{-\gamma_q^{} t}\right],\\
q_{+}^{(n-1)}(t,{\bf k}) &=& \int d\tau_{1}G_{r}(t,\tau_{1})\left(-\frac{2\gamma_q^{}}{L}\right)\theta(\tau_{1}-L) 
 \int d\tau_{2}G_{r}(\tau_{1},\tau_{2})\left(-\frac{2\gamma_q^{}}{L}\right)\theta(\tau_{2}-2L)\cdots\times\nonumber \\
 & & \int d\tau_{n}G_{r}(\tau_{n-1}-\tau_{n})\theta(\tau_{n}-(n-1)L) {\lambda_{q}\over M_q}
 f_{0+}\left(\tau_{n}-(n-1)L,\bar{\bf x}_q^{};{\bf k}\right),
\end{eqnarray}
with $M_{1}=1/[2(-\omega-i\gamma_q^{}+\Omega)]$ and $M_{2}=1/[2(-\omega-i\gamma_q^{}-\Omega)]$.

\subsection{Zeroth-Order Correlation Functions}
For the factorized 
initial state assumed in this paper with both the detector and the field being in their ground states
and uncorrelated, the elements of the
covariance matrix can be decomposed into two parts corresponding to the two sets
of operators in the mode decomposition,
namely,
\begin{eqnarray}
\langle\hat{Q}(t),\hat{Q}(t)\rangle  & = & \frac{1}{2\Omega_{r}}|q_{\rm a}(t)|^{2}+\int_{k_{3}>0}\frac{d^{3}k}{2\pi{}^{3}}\frac{1}{2\omega}|q_{+}(t,{\bf k})|^{2}\nonumber\\
& \equiv & \langle\hat{Q}(t),\hat{Q}(t)\rangle_{\rm a}+\langle\hat{Q}(t),\hat{Q}(t)\rangle_{\rm v}
\end{eqnarray}
Here `a' corresponds to the detector and `v' corresponds to the influence of the field's vacuum fluctuations.
$\langle\hat{P}(t),\hat{P}(t)\rangle$ and $\langle \hat{Q}(t),\hat{P}(t)\rangle$ are similar.

In the weak coupling limit, the effect of reflected influences correspond to terms in (\ref{qexpan}) and (\ref{pexpan})
which are of higher (than zeroth) order;
consequently, at early time (up to $t\approx 1/\gamma_q $) accumulated effect of reflected influences is always small.
Thus for the purpose of studying the early time behavior of
entanglement, we now
ignore the contribution of reflected influences and
restrict ourselves to the lowest order correlations,
which read

\begin{eqnarray}
  & & \langle\hat{Q}(t),\hat{Q}(t)\rangle_{\rm a}^{(0)}\equiv\frac{1}{2\Omega_{r}}\left|q_{\rm a}^{(h)}(t)\right|^{2}\nonumber\\
  & &= \frac{1}{2\Omega_{r}}\left|\frac{1}{2}\left(1+\frac{\Omega_{r}+i\gamma_q^{}}{\Omega}\right)
    e^{-\gamma_q^{} t-i\Omega    t}+\frac{1}{2}\left(1-\frac{\Omega_{r}+i\gamma_q^{}}{\Omega}\right)e^{-\gamma_q^{} t+i\Omega t}\right|^{2},\\
 & &\langle\hat{Q}(t),\hat{Q}(t)\rangle_{\rm v}^{(0)}\equiv
   \int_{k_{3}>0}\frac{d^{3}k}{2\pi{}^{3}}\frac{1}{2\omega}\left|q_{+}^{(0)}(t,{\bf k})\right|^{2} \label{QQv}\\
 & &=\left(\frac{\lambda_{q}}{M_q\Omega}\right)^{2}
   \int_{k_{3}>0}\frac{d^{3}k}{2\pi{}^{3}}\frac{1}{2\omega}\left(\sin\frac{k_{3}L}{2}\right)^{2}\left|(M_{1}-M_{2})e^{-i\omega t}+
   (M_{2}e^{i\Omega t}-M_{1}e^{-i\Omega t})e^{-\gamma_q^{} t}\right|^{2}\nonumber \\
 & &=\left(\frac{\lambda_{q}}{M_q\Omega}\right)^{2}\frac{1}{4\pi^{2}}
   \int d\omega {\omega\over 2}\left( 1-\frac{\sin\omega L}{\omega L}\right)
   \left|(M_{1}-M_{2})e^{-i\omega t}+(M_{2}e^{i\Omega t}-M_{1}e^{-i\Omega t})e^{-\gamma_q^{} t}\right|^{2},\nonumber
\end{eqnarray}
and so on.

Notice that the a-parts of the zeroth order correlators $\langle...\rangle_{\rm a}^{(0)}$ do not
depend on the distance $L/2$ between the detector and the boundary, but the
v-parts induced by vacuum fluctuations do have such dependence.

In \cite{LH_2IHO}, for the case of two inertial detectors distance $L$ apart,
which are located in free space with identical couplings to the field, the
zeroth order correlators due to vacuum fluctuations are given as,
for example
($M_q \equiv 1$ and $d$ in \cite{LH_2IHO} is replaced by $L$),
\begin{eqnarray}
 \langle\hat{Q}_{A}(t),\hat{Q}_{B}(t)\rangle_{\rm v}^{(0)}&=&\left(\frac{\lambda_{q}}{\Omega}\right)^{2}\frac{1}{4\pi^{2}}
 \int d\omega \,\omega\, \frac{\sin\omega L}{\omega L} \times \nonumber\\ & &
 \left|(M_{1}-M_{2})e^{-i\omega t}+\left(M_{2}e^{i\Omega t}-M_{1}e^{-i\Omega t}\right)e^{-\gamma_q^{} t}\right|^{2}.\label{QAQBv} 
\end{eqnarray}

Physically, the v-parts of the zeroth order correlators $\langle...\rangle^{(0)}_{\rm v}$
effectively measure the response of the detector to vacuum fluctuations of the field.
The similarity between the integrands of $\langle\hat{Q}(t),\hat{Q}(t)\rangle_{\rm v}$ in Eq.$(\ref{QQv})$ and
$\langle\hat{Q}_{A}(t),\hat{Q}_{B}(t)\rangle_{\rm v}$ in Eq.$(\ref{QAQBv})$ is not surprising.
In Appendix \ref{w2IHO} we show that, if we have
two inertial detectors C and D in free space at a distance $L$ apart,
with coupling constants of the same magnitude but in opposite signs,
then $(\hat{Q}_{C}(t)+\hat{Q}_{D}(t))/2$ obeys the same equation of motion
as the one for $\hat{Q}(t)$ in our model. Thus we see that
the self correlator $\langle\hat{Q}(t),\hat{Q}(t)\rangle_{\rm v}$ here
has the same value as $(\langle\hat{Q}_{C},\hat{Q}_{C}\rangle+\langle\hat{Q}_{D},\hat{Q}_{D}\rangle)/4
+\langle\hat{Q}_{C},\hat{Q}_{D}\rangle/2$ at $t$ and
contains the part of correlations of vacuum fluctuations in free space
which is odd with respect to the $z_3 =0$ plane, whereas
$\langle\hat{Q}_{C}(t),\hat{Q}_{D}(t)\rangle_{\rm v} \propto (-\lambda_q)\lambda_q$ here has exactly the same
value of $- \langle\hat{Q}_{A}(t),\hat{Q}_{B}(t)\rangle_{\rm v} $ in \cite{LH_2IHO}
because there the two detectors are identically coupled to the field
and so $\langle\hat{Q}_{A},\hat{Q}_{B}\rangle_{\rm v} \propto \lambda_q\lambda_q$.

\subsection{Early-time Dynamics of Detector-field Entanglement}

\begin{figure}[htbp]
  \center
    \includegraphics[width=0.54\textwidth]{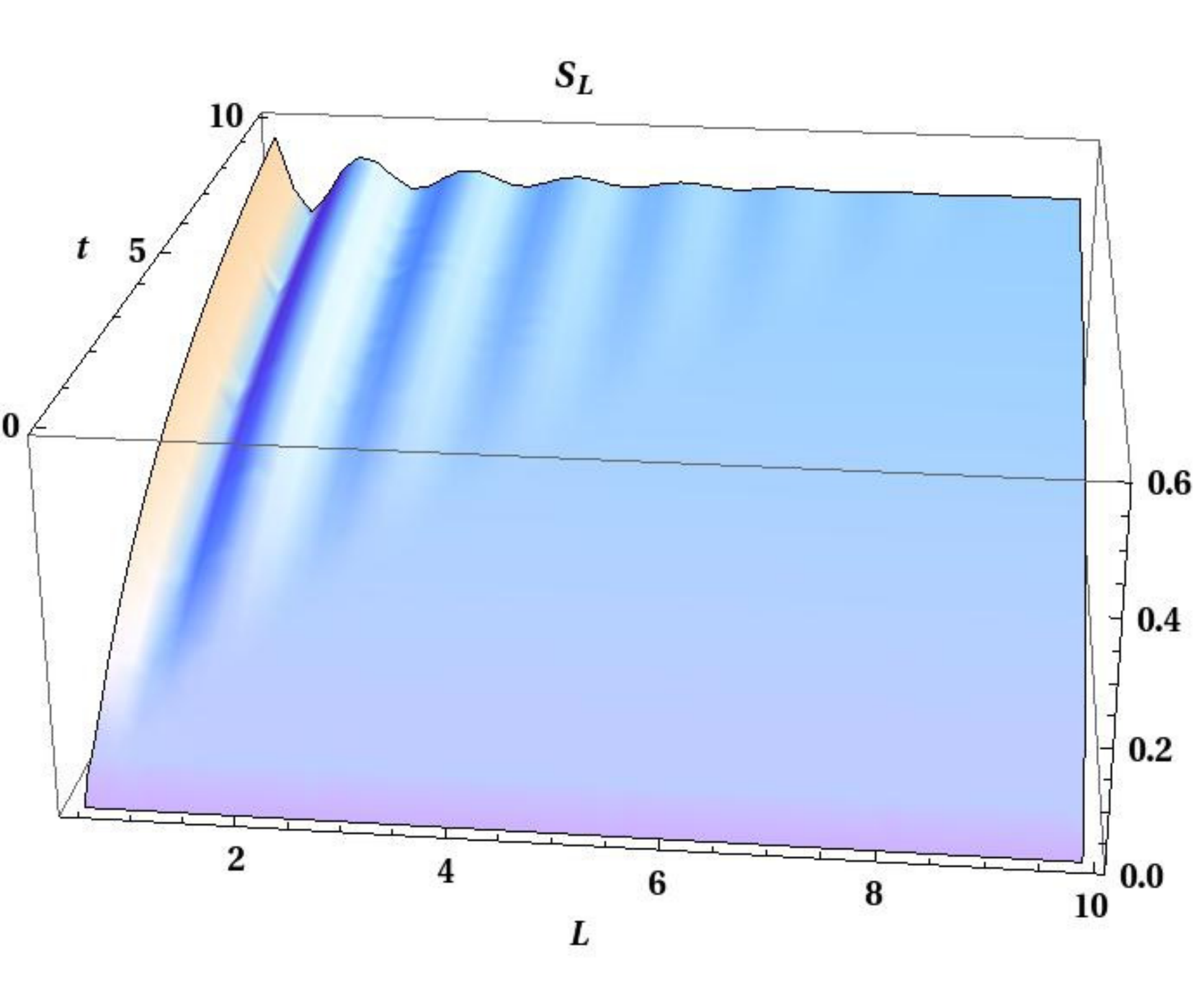}
    \includegraphics[width=0.44\textwidth]{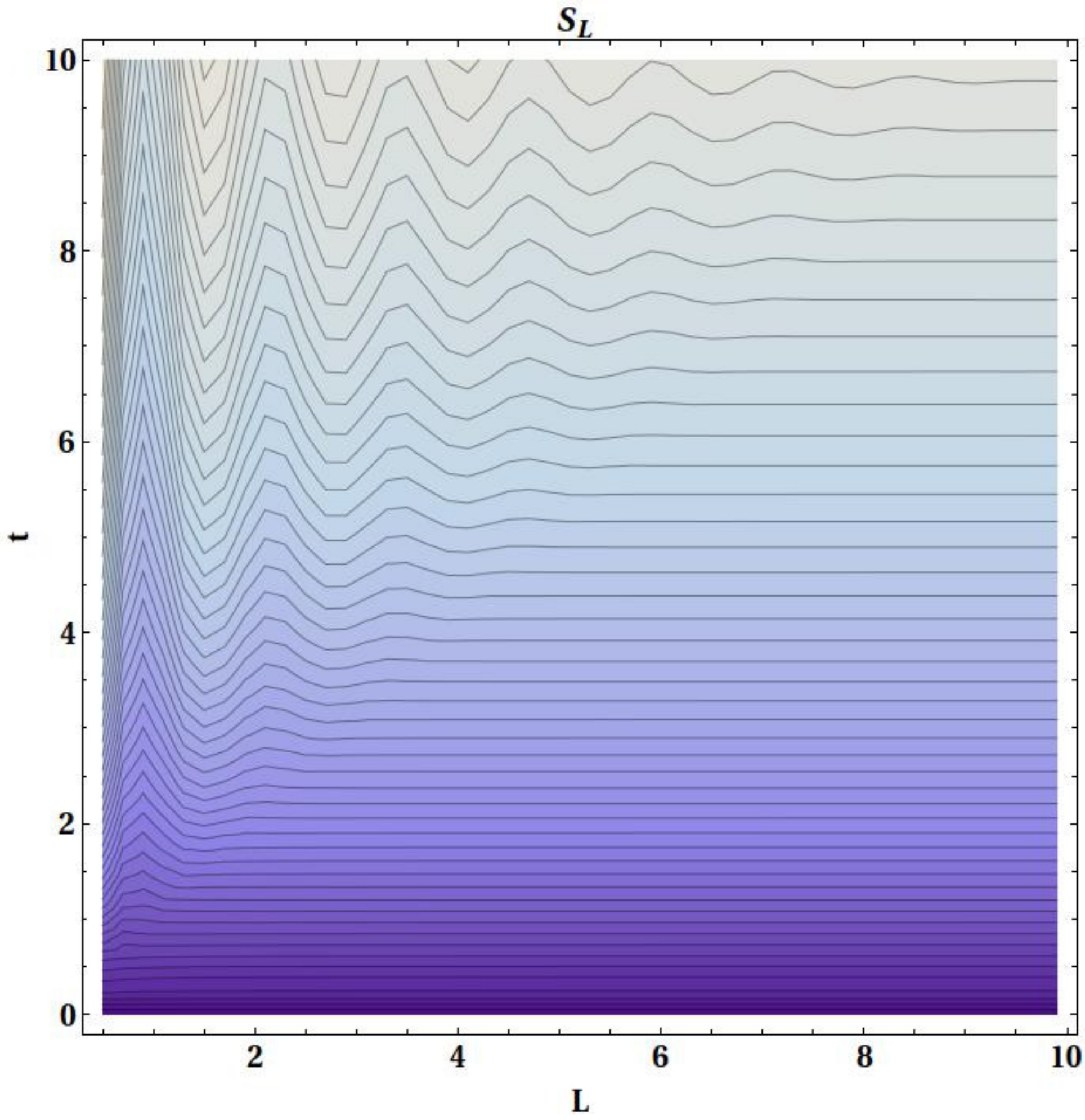}\\ 
    \includegraphics[width=0.49\textwidth]{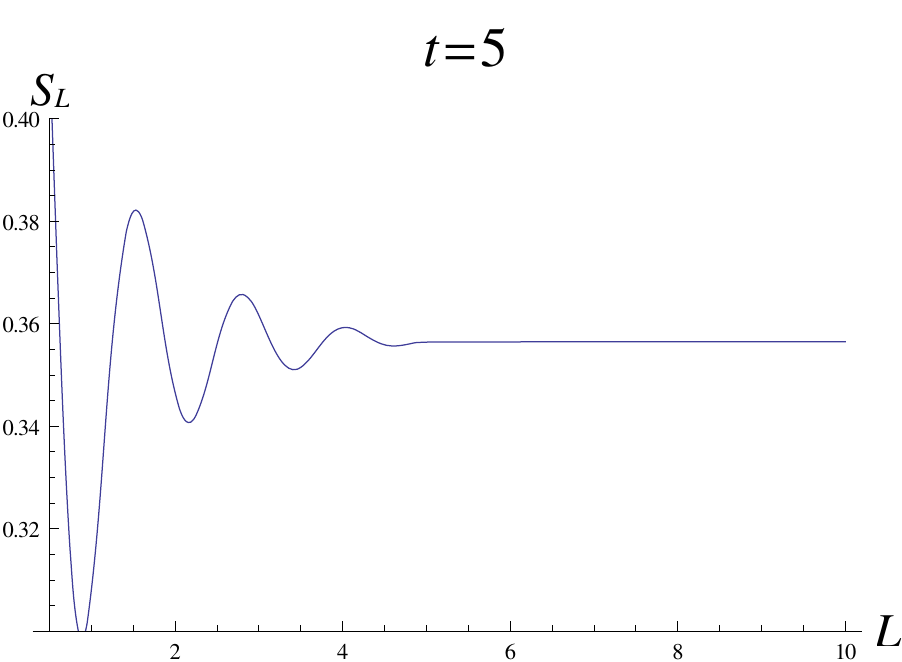}
    \includegraphics[width=0.49\textwidth]{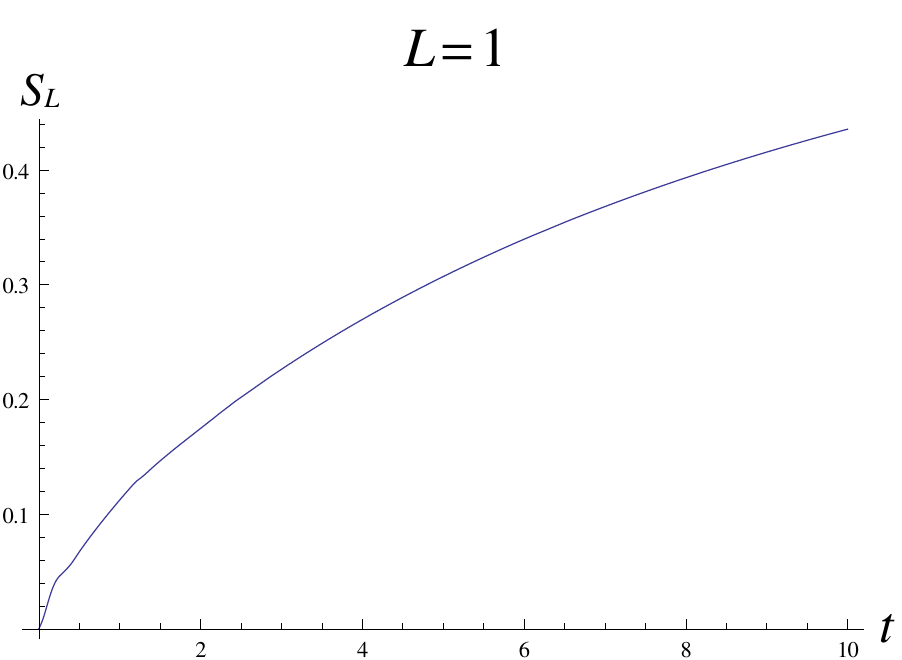}
     \caption{
	    Entanglement dynamics at early time where only zeroth order correlations
	    contribute.
	    Here $\gamma_q^{}=0.02$ and $\Omega=5$. The upper left plot shows the linear entropy as a function of L and t, with
	    the upper-right plot being its contour plot.
	    The lower-left plot shows the dependence of linear entropy on L at a given instant of time whereas the lower right
	    plot exhibits how the linear entropy evolves with time for the detector located at a certain distance.
	      }
\label{EntDynamics}
\end{figure}

With the previous results we can now demonstrate the evolution of detector-field
entanglement as the distance between the detector and the boundary changes.
Here we study the dependence of
the linear entropy on $L$ and $t$ numerically, as shown in Figure \ref{EntDynamics}, and provide physical explanation for its behavior.

Our numerical results reveal the following behaviors:
\begin{itemize}
\item For a given $L$, the entanglement between the detector and the field
increases monotonically at early times,
showing that the detector is getting
more and more entangled with the field after the interaction is turned on.
(See Figure \ref{EntDynamics} (upper row) and (lower right).)
\item Within the light cone,
at every given time in this stage,
the detector-field entanglement
exhibits an oscillatory behavior with spatial frequency being $\Omega$.
This shows that the growth rate of $S_L$
with which the detector gets entangled
with the field also oscillates as a function of $L$, as shown in
Figure \ref{EntDynamics} (upper row) and (lower left).
In \cite{Langlois} the transition rate of a detector close to an infinite plane boundary with Neumann boundary condition was shown
to exhibit similar oscillations.
\end{itemize}

Such oscillatory behavior in $L$ is solely due to the oscillatory behavior
of the zeroth order correlators corresponding to $\langle \hat{R}_{C}, \hat{R}_{D}\rangle_{\rm v}$,
$R=Q, P$, given in the last paragraph of the previous subsection.
After $\cos\theta$ in $k_3\equiv \omega\cos\theta$ has been integrated over the interval $[0, 1]$ in the integration
corresponding to $\langle\hat{Q}_{C},\hat{Q}_{D}\rangle_{\rm v}$ in  $(\ref{QAQBv})$,
the field modes in resonance with the detector have constructive (destructive) interference
at the local maxima (minima) of $S_L$ against $L$ at fixed $t$ at early times.

In \cite{LH_2IHO} the early time dynamics of entanglement between the two detectors exhibit
similar oscillatory dependence on the distance between them, which
is also due to distance-dependent correlations of vacuum fluctuations
experienced by the detectors.
More precisely, from $(\ref{QAQBv})$ one can see that the field modes with $\omega = n\pi/L$, $n=1,2,3,\cdots$,
has no contribution at all to the integration of $\langle\hat{Q}_{A},\hat{Q}_{B}\rangle_{\rm v}$, and those
satisfying $\tan \omega L = \omega L$ give the maximum and minimum values of the factor
$\sin \omega L/ (\omega L)$ in the integrand. In the weak coupling limit the integration in
$(\ref{QAQBv})$ is mainly contributed by the poles at $\omega \approx \pm \Omega$, namely,
those mode on resonance with the detectors. So $\langle\hat{Q}_{C},\hat{Q}_{D}\rangle_{\rm v} \approx 0$
for $L \approx n \pi/\Omega$, while $\langle\hat{Q}_{C},\hat{Q}_{D}\rangle_{\rm v}$ has local maximum or
minimum values when $L$ is about the solution of $\tan \Omega L = \Omega L$ in the weak coupling limit.

\section{Late-Time Stationary Limit of Detector-Field Entanglement}
\label{latetime}

In the late-time limit, as information flows from the detector to the field and gets dissipated through the
interaction with the field, the entire system approaches an asymptotically stationary state.
Under sufficiently general conditions, one can say that, supposing the field is initially in the vacuum, the late-time asymptotic
detector-field entanglement is independent of the initial condition of the detector as long as the detector-field coupling is linear.
As our result will show, the detector-field entanglement is a delicate manifestation of quantum correlations between the detector
and the field in the presence of the mirror.

\subsection{Quantum Langevin Equation and Covariance Matrix at Late-times}

One can compare the dynamics of the detector under the influence of the field
to the well-studied quantum Brownian motion (QBM) model, namely,
the retarded Green's functions and Hadamard elementary functions correspond to the dissipation
and noise kernels respectively and the function $I(\omega;{\bf x},{\bf y})$ as in (\ref{SpectralDensity})
represents the spectral density {(}see, e.g., \cite{HPZ,HM94} and
the references therein.{)} In the same vein Eq. (\ref{FI-AtomEqn})
can equivalently be written as a quantum Langevin equation
\begin{equation}
  M_{q}\ddot{\hat{Q}}(t)+M_{q}\Omega^{2}\hat{Q}(t)-\int_{t_{i}}^{t}d\tau\mu(t,\tau)\cdot\hat{Q}(\tau)=\hat{\xi}(t),\label{QBM-type-Eqn}
\end{equation}
in which the dissipation kernel
\begin{equation}
  \mu(t,s)=\lambda_{q}^{2}\ \theta(t-s)\int_{0}^{\Lambda}d\omega I(\omega;{\bf x}_q^{},{\bf x}_q^{})\sin\omega(t-s),
\label{eq:dissipationkernel}
\end{equation}
and the operator-valued stochastic force
$\hat{\xi}(t) = \lambda_{q}{\hat{\Phi}}_{0}({\bf x}_q^{},t)$  in the QBM language.

In the integrals over the frequency above and below we have assumed a
high frequency finite cutoff $\Lambda$ for the quantum field
which regularizes the quantum field's retarded Green's function from
which we obtain the effective equations of motion of the detector \cite{LH_2IHO}.

The noise kernel $\nu(t,s)$ 
quantifies the two-time correlation
of the Langevin forcing term $\hat{\xi}(t)$,
\begin{equation}
  \nu(t,s)\equiv\langle\{\hat{\xi}(t),\hat{\xi}(s)\}\rangle=\lambda_{q}^{2}\int_{0}^{\Lambda}d\omega
    I(\omega;{\bf x}_q^{},{\bf x}_q^{})\cos\omega(t-s).\label{eq:noisekernel}
\end{equation}
{[}From now on we simply denote $I(\omega)\triangleq I(\omega;{\bf x}_q^{},{\bf x}_q^{})$.{]}
Here the average is taken with respect to the initial state density matrix of the field.

By introducing the damping kernel $\gamma(t,s)$ defined by
\begin{equation}
\mu(\tau,s) \equiv M_{q}\frac{\partial}{\partial\tau}\gamma(\tau,s)=M_{q}\frac{\partial}{\partial s}\gamma(\tau,s),\label{eq:dampingkernel}\end{equation}
 we can bring (\ref{FI-AtomEqn}) to the final form
\begin{equation}
M_{q}\ddot{\hat{Q}}(t)+M_{q}\Omega_{r}^{2}\hat{Q}(t)+M_{q}\int_{t_{i}}^{t}d\tau
\gamma(t,\tau)\cdot\dot{\hat{Q}}(\tau)+2M_{q}\gamma(t,t_{i})\cdot\hat{Q}(t_{i})=
\hat{\xi}(t),\label{eq:Langevin}
\end{equation}
where $\Omega_{r}^{2}=\Omega^{2}+ \gamma_q^2$ is the renormalized frequency.

The general solution to (\ref{QBM-type-Eqn}) is given by
\begin{equation}
  \hat{Q}(t)=\hat{Q}_{0}(t)+\int_{t_{i}}^{t}ds\ \tilde{G}(t,s)\hat{\xi}(s),\label{Soln}\end{equation}
 where $\tilde{G}(t,s)$ is the retarded Green's function for (\ref{QBM-type-Eqn}) satisfying
\begin{equation}
  M_{q}\ddot{\tilde{G}}(t,s)+M_{q}\Omega^{2}\tilde{G}(t,s)-\int_{t_{i}}^{t}d\tau\mu(t,\tau)\cdot\tilde{G}(\tau,s)=\delta(t-s).\label{GreenF}
\end{equation}
For here and below all the Fourier components are defined for positive
frequency only, which corresponds to Fourier transformation of functions
defined on $t>0$. Accordingly, the Fourier space representation of
the above Green's function can be written as:
\begin{eqnarray}
  \tilde{G}(\omega) &\equiv& \left(M_{q}\left(-\omega^{2}+\Omega^{2}\right)-\mu(\omega)\right)^{-1}\label{Gmu} \\
  &=& \left(-\omega^{2}-i\omega\gamma(\omega)+\Omega_{r}^{2}\right)^{-1}M_{q}^{-1}.
\label{Ggamma}
\end{eqnarray}
In the late-time stationary limit, because $\mu(t,s)$ leads to dissipation
of the detector's free motion $\hat{Q}_{0}(t)$, we see that
\begin{equation}
  \tilde{\hat{Q}}(\omega)\to\tilde{G}(\omega)\tilde{\hat{\xi}}(\omega).
\end{equation}
So the elements of the covariance matrix at late times can be written as 
\begin{eqnarray}
  V_{QQ}^{\infty} &=&\left. \langle \hat{Q}(t),\hat{Q}(t) \rangle \right|_{t\to\infty} =
    \int_{0}^{\Lambda}d\omega\ \tilde{G}^{*}(\omega)\cdot I(\omega)\cdot\tilde{G}(\omega),\\
  V_{PP}^{\infty} &=&\left. \langle \hat{P}(t),\hat{P}(t) \rangle \right|_{t\to\infty}=
    \int_{0}^{\Lambda} d\omega\ \omega^{2}M_{q}^{2}\tilde{G}^{*}(\omega)\cdot I(\omega)\cdot\tilde{G}(\omega).
\end{eqnarray}
and $V_{QP}^{\infty}$ vanishes, as can be inferred from $V_{QP}(t)=M\dot{V}_{QQ}(t)/2$ while
$V_{QQ}(t)$ approaches an asymptotic constant value.
By applying the fluctuation-dissipation theorem \cite{CW51, Eck82, Eck84, FeynVern,HPZ}, 
one can eliminate the noise kernel in the covariance matrix elements, giving
\begin{eqnarray}
V_{QQ}^{\infty} &=& \frac{1}{\pi}\int_{0}^\Lambda 
    d\omega{\rm Im}[\tilde{G}(\omega)],\label{VQQinf}\\
V_{PP}^{\infty} &=& \frac{1}{\pi}\int_{0}^\Lambda 
    d\omega\ \omega^{2}M_{q}^{2}{\rm Im}[\tilde{G}(\omega)].\label{VPPinf}
\end{eqnarray}
For detailed derivation leading to the above results, please refer
to Appendix \ref{LateCV}.

\subsection{Entanglement between Detector and Field in Free Space}

Before we compute the late-time detector-field entanglement with a mirror present,
let us review the late-time detector-field entanglement
for a detector in free space for comparison.

The first step is to regularize the retarded Green's function of the field at the
trajectory of the detector. For a detector interacting with a massless
scalar field in free space, the entire system is governed by the same action
as Eq. (\ref{eq:Action_Half}) but without the $x_{3}>0$ restriction. 
The Dirichlet boundary conditions on the field in the z=0 plane are not needed here.

Following \cite{LH_2IHO} we obtain
\begin{equation}
\left(\partial_{t}^{2}+2\gamma_{q}^{}\partial_{t}+\Omega_{r}^{2}\right)\hat{Q}(t)=\lambda_{q}\hat{\Phi}(t,{\bf x}_q^{})/M_{q},\end{equation}
where $\Omega_{r}$ is the renormalized natural frequency which depends on cutoff $\Lambda$ and $\gamma_q \equiv \lambda_q^2/ 8\pi$.
For free space the retarded Green's function for (\ref{QBM-type-Eqn}) is given by
\begin{equation}
\tilde{G}(\omega)=\frac{1}{M_{q}}[-(\omega^{2}+i\gamma_q^{})^{2}+\tilde{\Omega}_{r}^{2}]^{-1},\end{equation}
 where $\tilde{\Omega}_{r}\equiv\Omega_{r}^{2}-\gamma^{2}$ from which
the late-time covariances can be computed.

In the weak coupling limit, up to the first order of $\gamma_q^{}$, one has
\begin{equation}
V_{QQ,{\rm free}}^{\infty}=\frac{1}{2M_{q}\tilde{\Omega}_{r}}\left(1-\frac{2\gamma_q^{}}{\pi\tilde{\Omega}_{r}}\right),\end{equation}
\begin{equation}
V_{PP,{\rm free}}^{\infty}=M_{q}\left(\frac{\tilde{\Omega}_{r}}{2}+\frac{1}{\pi}\gamma_q^{}\left[2(\ln\Lambda-\ln\tilde{\Omega}_{r})-\frac{\tilde{\Omega}_{r}^{2}}{\Lambda^{2}}-1\right]\right),
\end{equation}
and the late-time detector-field entanglement in free space is, to the same order of perturbation in $\gamma_q$,
\begin{equation}
 S_{L,\,{\rm free}}= \frac{\gamma_q^{}}{\pi \Omega_{r}}\left[2(\ln\Lambda-\ln\tilde{\Omega}_{r})-\frac{\tilde{\Omega}_{r}^{2}}{\Lambda^{2}}-2\right]
\end{equation}
Notice that there is a logarithmic dependence on cutoff scale $\Lambda$, because physically $\Lambda$ sets a restriction on
the ultraviolet modes that the detector can be entangled with.

\subsection{Entanglement between Detector and Field under Mirror Reflection}

In the presence of a perfect mirror, the entire system is governed
by action (\ref{eq:Action_Half}). The Heisenberg equation of motion
for the detector after the same regularization as above is
\begin{equation}
\left(\partial_{t}^{2}+2\gamma_q^{}\partial_{t}+\Omega_{r}^{2}\right)\hat{Q}(t)=-\frac{2\gamma_q^{}}{4\pi L}\theta(t-L)\hat{Q}(t-L)+\lambda_{q}\hat{\Phi}(t,{\bf x}_q^{}),\label{RegEOM_Half}
\end{equation}
[Cf.(\ref{EOMq+})]
and correspondingly\begin{equation}
\tilde{G}(\omega)=\frac{1}{M_{q}}[-(\omega+i\gamma_q^{})^{2}+\tilde{\Omega}_{r}^{2}+(2\gamma_q^{}e^{i\omega L}/L)]^{-1}.
\end{equation}
Here the last term inside the square brackets shows the difference
from the free space results. It can be interpreted as the contribution
from the detector's mirror image located at a vertical distance $L/2$ behind the mirror.

For this case the exact late-time covariance matrix becomes
\begin{eqnarray}
V_{QQ,{\rm half\,space}}^{\infty} & = & \frac{1}{\pi M_{q}}\int_{0}^{\Lambda}d\omega{\rm Im}\left[
\frac{1}{-(\omega+i\gamma_q^{})^{2}+\tilde{\Omega}_{r}^{2}+(2\gamma_q^{}e^{i\omega L}/L)}\right],\label{eq:Vxx} \\
V_{PP,{\rm half\,space}}^{\infty} & = & \frac{M_{q}}{\pi}\int_{0}^{\Lambda}d\omega{\rm Im}\left[
\frac{\omega^{2}}{-(\omega+i\gamma_q^{})^{2}+\tilde{\Omega}_{r}^{2}+(2\gamma_q^{}e^{i\omega L}/L)}\right].\label{eq:Vpp}\end{eqnarray}
Assuming that the detector is only weakly coupled to the field, we can
perturbatively expand the above integrals and get
\begin{eqnarray}
V_{QQ,{\rm half\,space}}^{\infty} & = & V_{QQ,{\rm free}}^{\infty} +\delta V_{QQ}^{\infty}+
O(\gamma_{q}^{2}),\\
V_{PP,{\rm half\,space}}^{\infty} & = & V_{PP,{\rm free}}^{\infty} + \delta V_{PP}^{\infty}+O(\gamma_{q}^{2}),
\end{eqnarray}
where the terms $\delta V_{QQ}^{\infty}$ and $\delta V_{PP}^{\infty}$
represent the first corrections to the covariance matrix elements
due to the presence of the mirror. Physically keeping only these terms
in this perturbative expansion is equivalent
to ignoring the multiple reflections between the detector and the mirror.
The exact form for the leading order correction is given below:
\begin{align}
\delta V_{QQ}^{\infty} & = \frac{1}{\pi M_{q}}\int_{0}^{\Lambda}d\omega
{\rm Im}\left[\frac{1}{-(\omega+i\gamma_q^{})^{2}+\tilde{\Omega}_{r}^{2}}\left(\frac{-2\gamma_q^{}e^{i\omega L}/L}{-(\omega+i\gamma_q^{})^{2}+\tilde{\Omega}_{r}^{2}}\right)\right]\nonumber\\
 & =-\frac{1}{\pi}\frac{1}{M_{q}\Omega_{r}}\frac{\gamma_q^{}}{L}{\rm Re}\left[\left(i\frac{1}{\Omega_{r}^{2}}+\frac{L}{\Omega_{r}}\right)e^{i\Omega_{r}L}\Gamma[0,i\Omega_{r}L]\right],
 \label{dVxx}
\end{align}
\begin{eqnarray}
\delta V_{PP}^{\infty} & = & \frac{M_{q}}{\pi}\int_{0}^{\Lambda}d\omega {\rm Im}\left[\frac{\omega^{2}}{-(\omega+i\gamma_q^{})^{2}+\tilde{\Omega}_{r}^{2}}\left(\frac{-2\gamma_q^{}e^{i\omega L}/L}{-(\omega+i\gamma_q^{})^{2}+\tilde{\Omega}_{r}^{2}}\right)\right]\nonumber\\
 & = & -\frac{M_{q}^{}\gamma_q^{}}{\pi\Omega_{r}L}{\rm Re}\left[\left(-i+L\Omega_{r})\right)
 e^{i\Omega_{r}L}\Gamma[0,i\Omega_{r}L]\right],
\label{dVpp} 
\end{eqnarray}
in the limit of large cutoff $\Lambda$.

The change of linear entropy due to the presence of the mirror, compared
to the case of free space, is given as
\begin{equation}
\Delta S_{L}
\equiv S_{L,\,{\rm half\, space}} - S_{L,\,{\rm free}}
\approx -\frac{2}{\pi}\frac{\gamma_q^{}}{\Omega_{r}}{\rm Re}\left[e^{i\Omega_{r}L}\Gamma[0,i\Omega_{r}L]\right]\label{eq:LEntropy}
\end{equation}
up to the leading order.
Note that despite of the dependence of both $S_{L,\,{\rm half\, space}} $ and $S_{L,\,{\rm free}}$ 
on the ultraviolet cutoff $\Lambda$, the difference between them is insensitive to $\Lambda$.

\subsection{Behavior and Physical Interpretation of Late-time Detector-Field Entanglement}

\begin{figure}
\includegraphics[width=6cm]{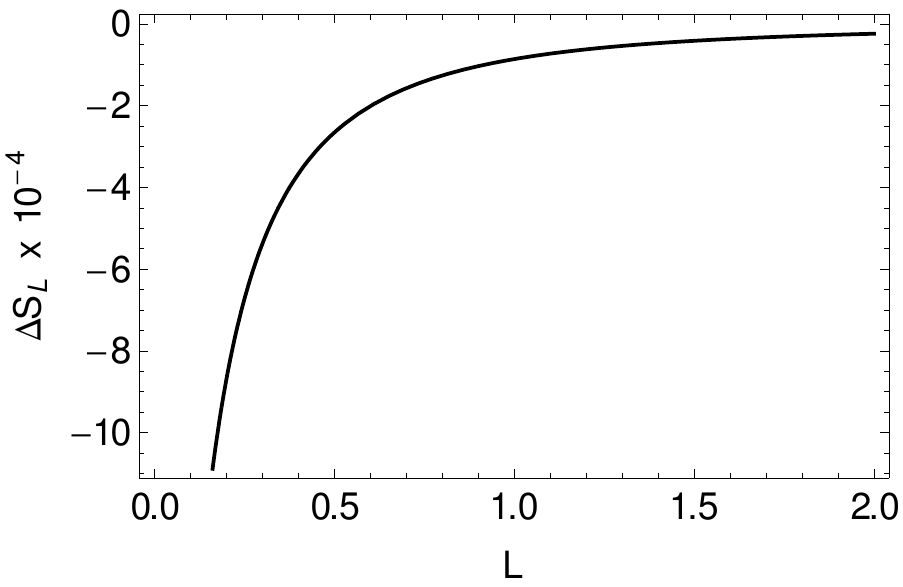}
\includegraphics[width=6cm]{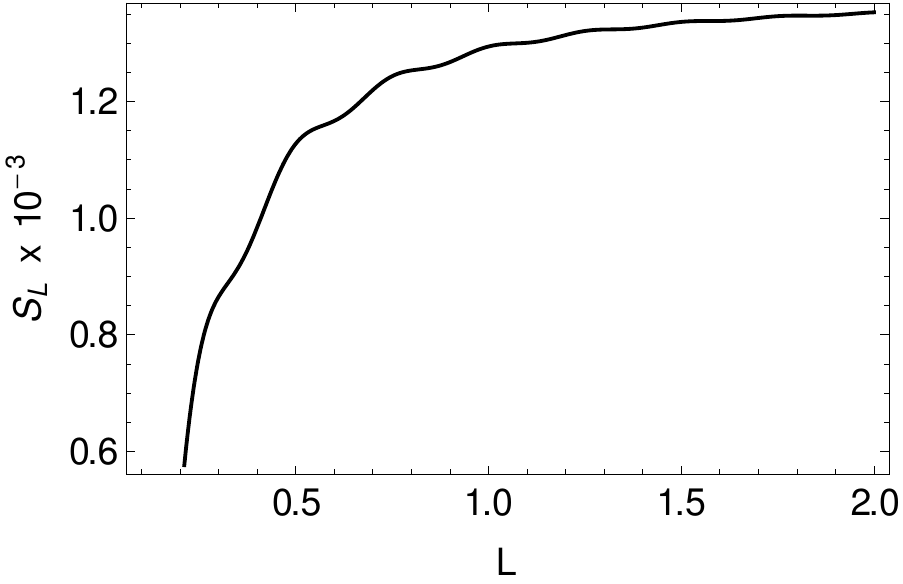} \\
\centering \includegraphics[width=6cm]{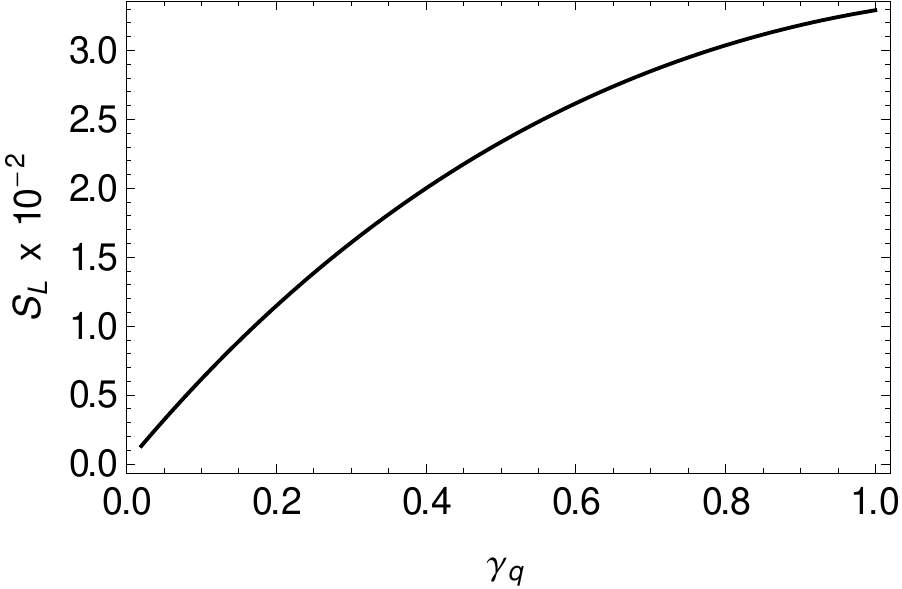}
\caption{(Upper-left) Leading-order corrections to linear entropy given in $(\ref{eq:LEntropy})$
as a function of distance between the
detector and its image due to the presence of the mirror.
(Upper-right) Numerical result of the linear entropy of the detector to all orders
as a function of distance between the detector
and its image according to Eqs (\ref{eq:Vxx}) and (\ref{eq:Vpp}).
(Lower) Linear entropy as a function of $\gamma_q$.
Here $M_{q}=1$, $\gamma_q^{}=0.02$, $\Omega_{r}=5$.}
\label{LinearEntropy}
\end{figure}

The leading order corrections $\Delta S_L$ given in $(\ref{eq:LEntropy})$ is negative
(see figure \ref{LinearEntropy} (upper-left)), indicating that
the presence of the mirror acts to reduce the linear entropy between the detector and the field,
thereby causing them to be less entangled. 
Furthermore, in figure \ref{LinearEntropy} (upper-right), we see that the numerical result of $S_{L,\,{\rm half\, space}}$ 
to all orders is monotonically decreasing, though wiggling, as $L$ decreases.
Clearly in the late-time stationary limit, the detector-field entanglement is suppressed in the presence of the mirror
compared to the case of free space. Such behavior can be 
understood as follows.

Since the combined detector-field system is in a global pure state, 
the detector-field entanglement roughly measures 
the strength of quantum correlation between the two parties.
Indeed, as the detector becomes more and more strongly coupled with the field, 
the correlation between the detector and the field get stronger, and so
the detector-field entanglement increases monotonically, 
as shown in the lower plot of Figure \ref{LinearEntropy}.

The Heisenberg equation of motion (\ref{RegEOM_Half}) for the detector's internal degree of freedom shows that, 
through field propagation in the form of spherical wave reflected by the mirror, 
the detector's internal degree of freedom will get a retarded influence from 
itself after time L. 
There the negative sign attached to the reflected component comes from the Dirichlet boundary condition enforced by the mirror. 
This condition acts on the detector as a time-delayed negative feedback.
As a result, the correlation established
between the detector and the field arising from their interactions will be effectively reduced, causing the detector to be less entangled
with the field. At a closer separation  between the detector and the mirror, the magnitude of the reflected negative influence becomes
larger, causing even stronger suppression of the quantum entanglement between the detector and the field.

\subsection{Effect of Mirror Image versus Effect of Real Object}

It is of interest to ask, in our model, whether the
effect of the mirror on late-time detector-field entanglement can be reproduced
by a real detector located at the position of the mirror image.
In the setup above, 
the lowest order correction to self-correlators corresponds
to the physical process in which the detector emits a quanta and then
interact with it after it is reflected by the mirror. This process
contributes a correction of order $O(\gamma_q^{}/L)$,
according to $(\ref{dVxx})$ and $(\ref{dVpp})$.
On the other hand,
in the cases 
of two inertial detectors with coupling
constant equal in magnitude but opposite in signs, as stated in Appendix \ref{w2IHO},
the lowest order correction to self-correlators corresponds
to the following physical process: detector A emits a quanta,
which interacts with detector B, then the back reaction
from detector B to the field echoes back and interacts with detector A.
The contribution of this process is of order $O(\gamma_q^{2}/L^{2})$.
Therefore, these two cases correspond to
two different physical processes involving different orders in the coupling constants.

\section{Discussion}
\label{discuss}
The goal of this paper is to probe into the effect of boundaries on the quantum entanglement 
between local objects and a quantum field, such as that between an atom and its trap or cavity surface.
Toward this goal we consider 
the model of an UD detector interacting linearly with a massless scalar quantum field under
the boundary condition of a perfect mirror.
Mathematically, the mirror alters the spectral density of the quantum field by imposing a Dirichlet boundary condition on it and
subsequently affects the dynamics of the detector. We describe the detector-field entanglement, both in its
early-time dynamics and its late-time stationary behavior. We show that in both cases the behavior can be depicted in terms of the influence of a  mirror image.

One can see more clearly which parties are being entangled if one considers a microscopic model of a mirror
 such as the one considered by Galley et al \cite{OM-T1}, where the mirror's internal degree of freedom is  modeled by an oscillator
(called mirosc) with very light mass (whereby the quantum fluctuations of the mirror's internal degree of freedom will be suppressed).
In this setup we could then consider three physical degrees of freedom: the detector, the field and the mirosc. If one first considers the interaction between the mirosc and the field, this would yield in the zero
mirosc mass  or infinite reflectivity limit the modified field configuration derived here. Then from the covariance matrix of the
detector one can derive the entanglement between the detector and the modified field. With a microphysical model of the mirror one
can calculate how the dynamics of the mirosc is altered while interacting with the field, how the field is modified, and how  the
detector is entangled with the modified field. The advantage of this approach is that one can see clearly how this entanglement can be interpreted
as the entanglement between the detector and the mirosc (see, e.g., \cite{BehuninHu} for an example of the successive levels of coarse-graining).
This reminds us of a similar procedure in electrostatics, namely, how the force between a charge and the induced surface charge density on
 a conducting plate can be calculated using the image charge method.

Another observation using elementary physics of wave reflection upon a mirror is  the following:
Since the field configuration is at the base of inquires into boundary effects on the detector field entanglement,
the behavior of reflected waves could provide some useful guide in building up our intuition on quantum entanglement in this setting.
Instead of a mirror with perfect reflectivity we can think of two
adjoining dielectric media 1, 2 with dielectric coefficients $\e_{1}<\e_{2}$
(the mirror situation considered above corresponds to the case where
$\e_{1}$ is the vacuum $\e_{0}$, much smaller than $\e_{2}$). For
waves propagating from a soft medium 1 to a hard medium 2, the reflected
wave is inverted. This shows up in the reflected field configuration
carrying an opposite sign from the original field configuration, thus
partially canceling it. This cancellation effect is more severe near
the mirror surface and hence we see the decrease of entanglement as
the detector gets closer to the mirror. If this reasoning is correct then
in the reverse situation, say a detector in water facing the sky (the air is medium 1, where we have
assumed $\e_{2}>\e_{1}$), as waves from the heavy
medium entering a light medium will be reflected at the interface
with a positive amplitude, the entanglement would increase as the
detector comes up close to the water-air surface. When the experimental techniques improve to the degree that one can
measure the quantum entanglement between an atom and the trap surface
these results could show their practical value.

A natural question to ask is what will be altered if we replace the perfectly reflecting mirror here 
by a more realistic dielectric medium. This is carried out in a sequel paper \cite{ZBLH2} which treats 
the quantum entanglement between an atom and a dielectric medium by means of the influence functional method.
As a small corollary from this work 
we will be able to check on the correctness of the above qualitative
argument based on symmetry and parity considerations. Later papers
in this series will address entanglement domain and entanglement pattern,
and a parallel series on quantum entanglement in topologically non-trivial
spaces starting with $R^{1}\times S^{1}$ \cite{ZCLH1}, which can be applied to atoms in a toroidal trap.\\

\noindent{\bf Acknowledgement} This work is supported in part by
the NSF Grant No. PHY-0801368 and the Nation Science Council of Taiwan under the Grant No. NSC 99-2112-M-018-001-MY3
and in part by the National Center for Theoretical Sciences, Taiwan.
ROB is aided by an NSF-NSC U.S.-East Asia Ph.D student grant award to spend a summer in Taiwan in 2010.

\appendix

\section{Derivation of late-time covariance matrix}
\label{LateCV}
According to our definition, the full-time, exact expression for the
QQ-element of late time covariance matrix is
\begin{align}
V_{QQ}(t) & =\int_{0}^{\infty}d\omega\cdot I(\omega)\int_{0}^{t}d\tau_{1}\int_{0}^{t}d\tau_{2}\tilde{G}(\tau_{1})\cos[\omega(\tau_{1}-\tau_{2})]\tilde{G}(\tau_{2})\\
 & =\int_{0}^{\infty}d\omega\cdot I(\omega)\int_{0}^{t}d\tau_{2}\int_{\tau_{2}}^{\tau_{2}+t}d\bar{\tau}\tilde{G}(\bar{\tau}-\tau_{2})
\cos[\omega(\bar{\tau}-2\tau_{2})]\tilde{G}(\tau_{2}).
\end{align}
At late times the integral can be approximated by
\begin{align}
V_{QQ}(t) & \approx\int_{0}^{\infty}d\omega\cdot I(\omega)\int_{0}^{t}d\tau_{2}\int_{\tau_{2}}^{t}d\bar{\tau}\tilde{G}(\bar{\tau}-\tau_{2})\cos[\omega(\bar{\tau}-2\tau_{2})]\tilde{G}(\tau_{2})\\
 & =\int_{0}^{\infty}d\omega\cdot I(\omega)\int_{0}^{t}d\bar{\tau}\int_{0}^{\bar{\tau}}d\tau_{2}\tilde{G}(\bar{\tau}-\tau_{2})\cos[\omega(\bar{\tau}-2\tau_{2})]\tilde{G}(\tau_{2})\\
 & =\int_{0}^{\infty}d\omega\cdot I(\omega)\int_{0}^{t}d\tau Re\left\{[e^{-i\omega\tau}\tilde{G}(\tau)]*[e^{i\omega\tau}\tilde{G}(\tau)]\right\}.\end{align}
Here $*$ denotes convolution.

When $t\rightarrow+\infty$, the above approximated expressions become
exact, and we have
\begin{equation}
V_{QQ}^{\infty}=\int d\omega\ \tilde{G}^{*}(\omega)\cdot I(\omega)\cdot\tilde{G}(\omega),\end{equation}
and similarly, the PP-element of the exact late-time covariance matrix is
\begin{equation}
V_{PP}^{\infty}=\int d\omega\ \omega^{2}M_{q}^{2}\tilde{G}^{*}(\omega)\cdot I(\omega)\cdot\tilde{G}(\omega).
\end{equation}
According to previous definitions (\ref{eq:dissipationkernel}) and (\ref{eq:dampingkernel})
we have
\begin{equation}
I(\omega)=\frac{2}{\pi}\omega M_{q}Re[\gamma(\omega)],
\end{equation}
which follows from the fluctuation-dissipation relation between noise and dissipation kernels, or equivalently their common 
relation with the spectral density. Then from (\ref{Ggamma}),
\begin{eqnarray}
& & \tilde{G}^{*}(\omega)\cdot Re[\gamma(\omega)]\cdot\tilde{G}(\omega) \nonumber\\
& &=\frac{1}{2}\left[\tilde{G}^{*}(\omega)\cdot\gamma^{*}(\omega)\cdot\tilde{G}(\omega)+
  \tilde{G}^{*}(\omega)\cdot \gamma(\omega)\cdot\tilde{G}(\omega)\right] \nonumber \\
& &=\frac{1}{2}\left[\left(1-(-\omega^{2}+\Omega_{r}^{2})\tilde{G}^{*}(\omega)M_{q}\right)
     \frac{\tilde{G}(\omega)}{2i\omega M_q}
   +\frac{\tilde{G}^{*}(\omega)}{(-2i\omega M_q)}\left(1-(-\omega^{2}+\Omega_{r}^{2})M_{q}\tilde{G}(\omega)\right)\right]\nonumber \\
& &=\frac{1}{2\omega M_q}Im[\tilde{G}(\omega)].
\end{eqnarray}
Therefore
in the late-time limit, we can express the covariance matrix elements as
(\ref{VQQinf}) and (\ref{VPPinf}) where the explicit reference to the noise kernel is eliminated
as $\tilde{G}(\omega)$ in (\ref{Gmu}) depends only on $\mu(\omega)$.

\section{Comparison with the case of two inertial detectors}
\label{w2IHO}

In Section VI of Ref. \cite{LH_2IHO}, we have obtained the late-time
correlators in the case with two identical UD detectors
at rest at $x_{3}=\pm L/2$. Let $\lambda_{0}\to-\lambda_{q}$
for the left detector ($Q_{A}$) and $\lambda_{0}\to+\lambda_{q}$
for the right detector ($Q_{B}$), one may wonder whether the detector
on the right ($Q_{B}$ at $x_{3}=+L/2$) in this two-detector case
would behave the same as the detector at the same position (say, $\tilde{Q}_{B}$
at $x_{3}=+L/2$) in the above single-detector case with its image
detector.

The answer is no. The presence of the other detector separated in
a distance $L$ from one detector introduces corrections to the late-time
correlators of a single detector, which are $O(\gamma_q^{2}/L^{2})$
for the self correlators and $O(\gamma_q^{}/L)$ for the cross correlators.
This is different from the above self correlators $V_{QQ}$ and $V_{PP}$,
which have $\delta V_{QQ}$ and $\delta V_{PP}$ in $O(\gamma_q^{}/L)$.

This can be understood as follows. The late-time behavior of the correlators
in \cite{LH_2IHO} are determined by mode functions $q_{A}^{(+)}(t,{\bf k})$
and $q_{B}^{(+)}(t,{\bf k})$, whose equation of motion reads (Eq.(13)
in \cite{LH_2IHO} with $d$ and $\lambda_{0}$ modified)
\begin{eqnarray}
\left(\partial_{t}^{2}+2\gamma_q^{}\partial_{t}+\Omega_{r}^{2}\right)q_{B}^{(+)}(t,{\bf k}) & = & -\frac{2\gamma_q^{}}{ L}\theta(t-L)q_{A}^{(+)}(t-L,{\bf k})+\lambda_{q}e^{-i\omega t+ik_{3}L/2},\label{eomqB}\\
\left(\partial_{t}^{2}+2\gamma_q^{}\partial_{t}+\Omega_{r}^{2}\right)q_{A}^{(+)}(t,{\bf k}) & = & -\frac{2\gamma_q^{}}{ L}\theta(t-L)q_{B}^{(+)}(t-L,{\bf k})-\lambda_{q}e^{-i\omega t-ik_{3}L/2}.\label{eomqA}\end{eqnarray}

Similar to Eqs. $(A6)$-$(A11)$ in \cite{LH_2IHO}, these equations
give the same $c_{{\bf k}}^{0}$ for $q_{B}^{(+)}$ like $(A11)$,
while the counterpart for $Q_{A}$ is $-c_{{\bf k}}^{0}$. So the
late-time self correlators are still given by Eqs. $(48)$ and $(50)$
in \cite{LH_2IHO}, and the cross correlators are those in Eqs. $(49)$
and $(51)$ multiplied by $-1$. From Eqs. $(54)$ and $(55)$ in
\cite{LH_2IHO}, one can see that in weak coupling limit the late-time
cross correlators are $O(\gamma_q^{}/L)$ and the correction to
the self correlators is $O(\gamma_q^{2}/L^{2})$.

On the other hand, the detector in this paper has
\begin{equation}
\left(\partial_{t}^{2}+2\gamma_q^{}\partial_{t}+\Omega_{r}^{2}\right)q_{+}(t,{\bf k})=
-\frac{2\gamma_q^{}}{ L}\theta(t-L)q_{+}(t-L,{\bf k})+{\lambda_{q}\over M_q}e^{-i\omega t}\sin\frac{k_{3}L}{2},\label{eomQ1B}
\end{equation}
where $q_{+}(t-L,{\bf k})$ on the right hand side can
be interpreted as the image of $q_{+}$. Indeed, $(\ref{eomQ1B})$ gives 
\begin{equation}
\left. q_{+}(t,{\bf k})\right|_{t\gg1/\gamma_q^{}}=\frac{-{\lambda_{q}}e^{-i\omega t}\sin\frac{k_{3}L}{2}/M_q}{\omega^{2}+2i\gamma_q^{}\omega-\Omega_{r}^{2}-(2\gamma_q^{}e^{i\omega L}/L)},
\end{equation}
so at late times,
\begin{eqnarray}
\langle \hat{Q}(t),\hat{Q}(t)\rangle_{\rm v}
&= & \hbar\int\frac{d^{3}k}{(2\pi)^{3}2\omega}\left| q_{+}(t,{\bf k})\right|^{2}\nonumber \\
 && \stackrel{t\to\infty}{\longrightarrow}
 \frac{\hbar}{\pi M_q}\int_{0}^{\infty}d\omega\,{\rm Im}\left[\frac{1}{\omega^{2}-2i\gamma_q^{}\omega-\Omega_{r}^{2}-
 ( 2\gamma_q^{} e^{-i\omega L}/L)}\right],
\end{eqnarray}
which is exactly the $V_{QQ}^{\infty}$ in $(\ref{eq:Vxx})$ after
letting $\hbar=1$. 

Note that the first order correction to $V_{QQ}^{\infty}$ in free
space has exactly the same value as the late-time cross correlator
$\langle \{\hat{Q}_{A},\hat{Q}_{B}\}\rangle$ in the case
with two inertial detectors considered above. However, the latter
will not enter the reduced density matrix of $\hat{Q}_{B}$. Compare
$(\ref{eomQ1B})$ with $(\ref{eomqB})$ and $(\ref{eomqA})$, one
can see that it is $(q_{B}^{(+)}+q_{A}^{(+)})/(2i)$ rather than $q_{B}^{(+)}$
has the same late-time behavior as $q_{+}$
for $M_q=1$.


\begin{thebibliography}{99}

\bibitem{Casimir} H. B. G. Casimir, {\it On the attraction between two perfectly conducting plates}, 
Proc. K. Ned. Akad. Wet. {\bf 51}, 793 (1948).

\bibitem{DavFul} P. C. W. Davies and S. A. Fulling, {\it Radiation from moving mirrors and from black
holes}, Proc. Roy. Soc. Lond. A {\bf 356}, 237 (1977).

\bibitem{Hawk75} S. W. Hawking, {\it Particle creation by black holes}, Comm. Math. Phys. {\bf 43}, 199
(1975).

\bibitem{Unr76} W. G. Unruh, {\it Notes on black-hole evaporation}, Phys. Rev. D {\bf 14}, 870 (1976).

\bibitem{QFTBounTop} B. S. DeWitt, C. F. Hart, and C. J. Isham, in 
{\it Themes in Contemporary Physics} ed S. Deser (North Holland, Amsterdam, 1979)

\bibitem{QFText}{\it Proceedings of the 9th Conference on Quantum Field Theory Under the Influence of External Conditions (QFEXT09)}, 
edited by K. A. Milton and M. Bordag (World Scientific, Singapore, 2010).


\bibitem{Schr35} E. Schr\"odinger, {\it Discussion of Probability Relations between Separated Systems}, Proc. Camb. Phil. Soc. {\bf 31}, 555 (1935).

\bibitem{DeW79} B. S. DeWitt, in {\it General Relativity: an Einstein Centenary Survey}, edited by S. W. Hawking and W. Israel 
(Cambridge Univ. Press, Cambridge, 1979).

\bibitem{RQI}  
Special Issue on Relativistic Quantum Information,
Class. Quantum Grav. {\bf 29} (2012). 

\bibitem{HLL}  B. L. Hu, S.-Y. Lin, and J. Louko, 
{\it Relativistic quantum information in detector-field interaction}, 
Class. Quantum Grav. {\bf 29}, 224005 (2012). 

\bibitem{LinHuOLCE} S.-Y. Lin and B. L. Hu, 
{\it Entanglement creation between two causally disconnected objects}, 
Phys. Rev. D {\bf 81}, 045019 (2010).

\bibitem{QIPspecial} B. L. Hu and T. Yu, eds, Special Issue on Quantum Decoherence and 
Quantum Entanglement in Quantum Information Processing (December 2009 issue). 

\bibitem{YuEberly} T. Yu and J. H. Eberly,
{\it Finite-Time Disentanglement Via Spontaneous Emission},
Phys. Rev. Lett \textbf{93}, 140404 (2004).

\bibitem{PazRoncaglia} J. P. Paz and A. J. Roncaglia,
{\it Dynamics of the Entanglement between Two Oscillators in the Same Environment},
Phys. Rev. Lett. \textbf{100}, 220401 (2008);
{\it Dynamical Phases for the Evolution of the Entanglement between Two Oscillators Coupled to the Same Environment},
Phys. Rev. A \textbf{79}, 032102 (2009).

\bibitem{SCH1} K. Sinha, N. Cummings, and B. L. Hu,
{\it Protecting and Dynamically Generating Entanglement in a Two-Atom Two-Field-Mode Model},
preprint [arXiv:1004.1834].

\bibitem{Sabrina} M. Scala, B. Militello, A. Messina, S. Maniscalco, J. Piilo, and K. Suominen,
{\it Cavity Losses for the Dissipative Jaynes-Cummings Hamiltonian beyond Rotating Wave Approximation},
J. of Phys. A \textbf{40}, 14527 (2007).

\bibitem{Miao} H. Miao, S. Danilishin, and Y. Chen,
{\it Universal Quantum Entanglement between an Oscillator and Continuous Fields},
Phys. Rev. A \textbf{81}, 052307 (2010).

\bibitem{Marshall} W. Marshall, C. Simon, R. Penrose, and D. Bouwmeester,
{\it Towards Quantum Superposition of a Mirror},
Phys. Rev. Lett. \textbf{91}, 130401 (2003).

\bibitem{ASH} C. Anastopoulos, S. Shresta, and B. L. Hu,
{\it Entanglement Dynamics of Two Atoms Interacting Through a Quantum Field},
Quantum Information Processing {\bf 8}, 549 (2009), 
a summary of 
preprint [quant-ph/0610007].



\bibitem{FCAH_2atom} C. H. Fleming, N. Cummings, C. Anastopoulos, and B. L. Hu,
{\it Non-Markovian Dynamics and Entanglement of Two-Level Atoms in a Common field},
J. Phys. A. \textbf{45}, 065301 (2012).

\bibitem{SCH2} K. Sinha, N. Cummings, and B. L. Hu,
{\it Effect of Interatomic Separation on Entanglement Dynamics in a Two-Atom Two-Mode Model},
J. Phys. B: At. Mol. Opt. Phys. \textbf{45} (2012) 035503.

\bibitem{LH_2IHO} S.-Y. Lin and B. L. Hu,
{\it Temporal and Spatial Dependence of Quantum Entanglement from Field Theory Perspective},
Phys. Rev. D \textbf{79}, 085020 (2009).

\bibitem{SLCH} E.g., S.-Y. Lin, K. Shiokawa, C.-H. Chou, and B. L. Hu, 
{\it Quantum Teleportation between Moving Detectors in a Quantum Field},
preprint [arXiv:1204.1525].

\bibitem{Purcell} E. M. Purcell,
{\it Spontaneous Emission Probabilities at Radio Frequencies},
Phys. Rev. \textbf{69}, 681 (1946).


\bibitem{LinHu08} S.-Y. Lin and B. L. Hu,
{\it Quantum entanglement, recoherence and information flow in a particle-field system:
implications for black hole information issue},
Class. Quant. Grav. \textbf{25}, 154004 (2008).


\bibitem{OneModePurity} G. Adesso, A. Serafini and F. Illuminati,
{\it Determination of Continuous Variable Entanglement by Purity Measurements},
Phys. Rev. Lett. \textbf{92}, 087901 (2004).

\bibitem{Langlois} P. Langlois,
{\it Causal Particle Detectors and Topology},
Ann. Phys. (N.Y.) \textbf{321}, 2027 (2006).


\bibitem{CW51} H. B. Callen, T. A. Welton, 
{\it Irreversibility and Generalized Noise},
Phys. Rev. {\bf 83} 34 (1951).

\bibitem{Eck82} W. Eckhardt,
{\it First and second fluctuation-dissipation-theorem in electromagnetic fluctuation theory},
Opt. Commun. {\bf 41}, 305 (1982).

\bibitem{Eck84} W. Eckhardt,
{\it Macroscopic theory of electromagnetic fluctuations and stationary radiative heat transfer},
Phys. Rev. A {\bf 29}, 1991 (1984).

\bibitem{FeynVern} R. P. Feynman and F. L. Vernon, Jr.,
{\it The Theory of a General Quantum System Interacting with a Linear Dissipative System},
Ann. Phys. (N.Y.) \textbf{24}, 118 (1963).

\bibitem{HPZ} B. L. Hu, J. P. Paz, and Y. Zhang,
{\it Quantum Brownian Motion in a General Environment I.
Exact Master Equation with Nonlocal Dissipation and Colored Noise},
Phys. Rev. D \textbf{45}, 2843 (1992).

\bibitem{HM94} B. L. Hu and A. Matacz,
{\it Quantum Brownian Motion in a Bath of Parametric Oscillators:
A Model for System-Field Interactions},
Phys. Rev.D \textbf{49}, 6612 (1994).

\bibitem{QOS} C. H. Fleming, B. L. Hu, and A. Roura,
{\it Exact Analytical Solutions to the Master Equation of Quantum Brownian Motion for a
General Environment}, Annals of Physics \textbf{326}, 1207 (2011).

\bibitem{OM-T1} C. R. Galley, R. O. Behunin, and B. L. Hu, {\it Theory of Optomechanics:
Oscillator-Field Model of Moving Mirrors}, preprint [arXiv:1204.2569].

\bibitem{BehuninHu} R. O. Behunin and B. L. Hu,
{\it Nonequilibrium Atom-Dielectric Forces Mediated by a Quantum Field}
Phys. Rev. A \textbf{84}, 012902 (2011).

\bibitem{ZBLH2} R. Zhou, R. O. Behunin, S.-Y. Lin, and B. L. Hu,
{\it Entanglement between Atoms and Dielectric Medium}, in preparation.

\bibitem{ZCLH1} R. Zhou, C.-H. Chou, S.-Y. Lin, and B. L. Hu, in preparation.


\end{thebibliography}



\end{document}